\documentclass[twoside,twocolumn,english]{revtex4}
\usepackage[T1]{fontenc}
\usepackage[latin1]{inputenc}
\usepackage{graphicx}

\makeatletter

\newcommand{\boldsymbol}[1]{\mbox{\boldmath $#1$}}

\usepackage{graphicx}

\usepackage{babel}
\makeatother
\begin{document}

\title{Signatures of Discontinuity in the Exchange-Correlation Energy Functional Derived from
the Subband Electronic Structure of Semiconductor Quantum Wells}

\author{S. Rigamonti and C. R. Proetto}

\affiliation{Centro At\'{o}mico Bariloche and Instituto Balseiro, \\
 8400 S.C. de Bariloche, R\'{\i}o Negro, Argentina}

\date{\today{}}

\begin{abstract}
The discontinuous character of the
exact exchange-correlation $(xc)$ energy functional of Density
Functional Theory is shown to arise naturally in the subband
spectra of semiconductor quantum wells. Using an
\emph{ab-initio} $xc$ functional, including
exchange exactly and correlation in an exact partial way, a
discontinuity appears in the $xc$ potential, each
time a  subband becomes slightly occupied.
Exchange and correlation give opposite contributions to the
discontinuity, with correlation overcoming
exchange. The jump in the intersubband energy
is in excellent agreement with experimental
data.
\end{abstract}
\maketitle Density Functional Theory (DFT) has become the standard
calculational tool in physics and quantum chemistry for the study
of atomic, molecular, and solid state systems. The theory is based in
the Hohenberg-Kohn theorems\cite{parr}, that place the
ground-state electron density as the basic variable 
and provides a variational principle
for its calculation. Kohn and Sham (KS) showed how the problem of
variational minimization for the density could be
exactly mapped to one of non-interacting particles in
an effective potential, which contains only one
non-trivial component: the exchange-correlation ($xc$)
contribution\cite{parr}. DFT, however, gives no clue on how to
proceed for its practical calculation. Naturally a lot of
attention has been devoted to the development of better $xc$
energy functionals; KS introduced the highly successful Local
Density Approximation (LDA), which is widely employed nowadays,
along with the improvements born from it (such as the Generalized
Gradient Approximation or GGA, meta-GGA, etc.)\cite{tao}. The work
described here is motivated by this fundamental need of better
approximations for the $xc$ energy functional\cite{mattsson},
using as "laboratory" to test the accuracy of the
approximations the subband electronic structure of the quasi
two-dimensional electron gases (\emph{2DEG}) formed at the
interface between two dissimilar semiconductors, such as GaAs and
AlGaAs. In this Letter we show that at the one-subband
$\rightarrow$ two-subband quantum well (QW) transition ($1S
\rightarrow 2S$), the $xc$ potential behaves discontinuously, with
exchange and correlation giving opposite contributions (i.e.,
competing) to the discontinuity. The intersubband
energy, which also jumps abruptly at the transition, is in
excellent agreement with experiments.

Our model system is a semiconductor modulation-doped QW grown
epitaxially, as shown in the upper inset of Fig. 1.
Assuming translational symmetry in the $(x-y)$ plane (area $A$),
and proposing accordingly a solution of the type
$\phi_{i\mathbf{k}\sigma}(\boldsymbol{\rho},z)=\text{e}^{i\mathbf{k}\cdot\boldsymbol{\rho}}\xi_{i}^{\sigma}(z)/\sqrt{A}$,
with $\boldsymbol{\rho}$ the in-plane coordinate, the zero
temperature ground-state electron density can be obtained by
solving a set of effective one-dimensional KS equations of the
form: 
\begin{equation}
\left[-\frac{1}{2}\frac{\partial^{2}}{\partial
z^{2}}+V_{KS}^{\sigma}(z)\right]\xi_{i}^{\sigma}(z)=\varepsilon_{i}^{\sigma}\xi_{i}^{\sigma}(z),\label{schro}
\end{equation}
 where effective atomic units have been used. $\xi_{i}^{\sigma}(z)$
is the wavefunction for electrons in subband $i$
($i=1,\,2,\,...$), spin $\sigma$ ($\sigma=\uparrow,\,\downarrow$),
and eigenvalue $\varepsilon_{i}^{\sigma}$. The local
(multiplicative) KS potential $V_{KS}^{\sigma}(z)$ is the sum of
several terms:
$V_{KS}^{\sigma}(z)=V_{ext}(z)+V_{H}(z)+V_{xc}^{\sigma}(z)$.
$V_{ext}(z)$ is given by the epitaxial potential plus an external
electric field. $V_{H}(z)$ is the Hartree
potential. Within DFT, $V_{xc}^{\sigma}(z)=A^{-1}\delta E_{xc}/\delta
n_{\sigma}(z)$. Departing from the main stream in most
applications of KS-DFT, our $xc$ energy functional is an explicit
functional of the whole set of $\varepsilon_{i}^{\sigma}$'s and
$\xi_{i}^{\sigma}$'s,
$E_{xc}=E_{xc}[\{\varepsilon_{i}^{\sigma},\xi_{i}^{\sigma}\}]$,
but an implicit (in general unknown) functional of the
spin-resolved density $n_{\sigma}(z)$. The zero-temperature $3D$
electron density is
$n_{\sigma}(z)=\sum_{\varepsilon_{i}^{\sigma}<\mu}(\mu-\varepsilon_{i}^{\sigma})|\xi_{i}^{\sigma}(z)|^{2}/2\pi$,
with 
$\mu$ the chemical potential. By assuming a paramagnetic
situation, we drop the spin index $\sigma$ from this point.

\begin{figure}
\includegraphics[%
  bb=50bp 260bp 525bp 760bp,
  clip,
  width=8cm,
  keepaspectratio]{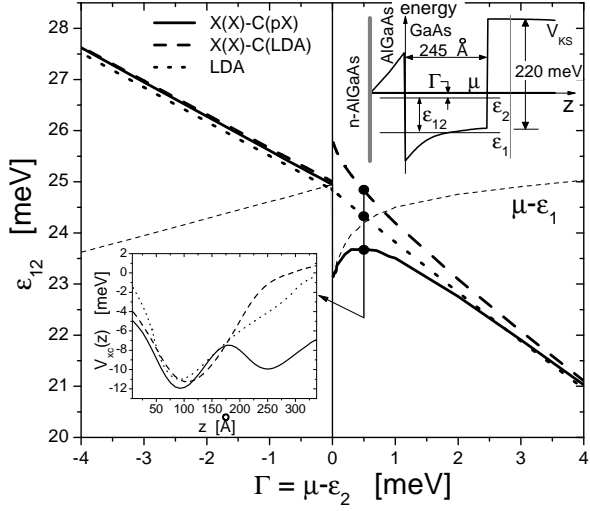}

\caption{Ground-state $\rightarrow$ first-excited intersubband
energy $\varepsilon_{12}$ as a function of $\Gamma$. Negative
(positive) values of $\Gamma$ correspond to the $1S$ ($2S$)
regime. Upper inset: schematic view of our model for the
modulation-doped QW. A charge-transfer field along $z$ is induced
by a distant metallic plate (thin vertical line at right). If the
metallic plate is positively (negatively) charged, more (less)
electrons are transferred towards the well from an ionized donor
impurities region (thick vertical stripe at left), which acts as a
particle reservoir fixing the chemical potential $\mu$. Lower
inset: $V_{xc}(z)$ in the same three different approximations as
for $\varepsilon_{12}$ ($\Gamma=0.5$ meV) . The QW extends from
$z=50$ \AA $\,\,\,$to $z=295 $ \AA.}
\end{figure}

The $xc$ energy functional $E_{xc}$ for our \emph{2DEG} has been
generated by G\"{o}rling-Levy (GL) perturbation theory where the
correlation energy is expanded in a series\cite{gorling},
\begin{equation}
E_{xc}[\{\varepsilon_{i},\xi_{i}\}]=E_{x}[\{\varepsilon_{i},\xi_{i}\}]+\sum_{n=2}^{\infty}E_{c}^{GL(n)}[\{\varepsilon_{i},\xi_{i}\}],\label{series}\end{equation}
 which is truncated at its leading contribution $n=2$. $E_{x}$ in
equation above is the exact exchange energy, known explicitly as a
Fock integral of the KS occupied orbitals. Its
multi-subband explicit expression is given by Eq.(38) in
Ref.\cite{rigamonti1} (denoted as I below). The terms
$E_{c}^{GL(n)}$ can be found explicitly\cite{gorling}. At
difference with $E_{x}$, however, they depend both on occupied and
an infinite number of unoccupied subbands. 
Their numerical evaluation is in consequence rather
expensive. The explicit expression for $E_{c}^{GL(2)}$ for our
semiconductor QW system is given by Eqs.(32) and (33) of I. A
similar correlation energy functional has been used in
Refs.\cite{mori} and \cite{jiang} for the case of atoms, with
mixed results. As shown here, Eq.(2) seems much more promising for
the \emph{2DEG}\cite{note}.

The next non-trivial problem is the evaluation of $V_{xc}(z)$,
given the already quoted implicit dependence of $E_{xc}$ on
$n(z)$ in Eq. (2). The procedure for dealing with implicit functionals relies on the
use of the chain rule for functional derivatives as follows \cite{gorling}:
\begin{equation} V_{xc}(z)=\frac{1}{A}\frac{\delta E_{xc}}{\delta
n(z)}=\frac{1}{A}\int\frac{\delta E_{xc}}{\delta
V_{KS}(z')}\frac{\delta V_{KS}(z')}{\delta
n(z)}dz'.\label{chain}\end{equation}
 To proceed with the calculation of $V_{xc}(z)$ \emph{directly} from Eq.(3),
we use a numerical method devised and explained in detail in I.
Eqs.(1)-(3) should be iterated until full self-consistency is
achieved.

We present in Fig. 1 the intersubband energy spacing
$\varepsilon_{12} \equiv \varepsilon_{2}-\varepsilon_{1}$, as a
function of $\Gamma \equiv \mu-\varepsilon_{2}$, for
three  approximations for
$E_{xc}$: LDA, exact-exchange plus
LDA for correlation [X(X)-C(LDA)], and X(X) plus partial
exact-correlation ($E_{c}^{GL(2)}$), denoted as X(X)-C(pX).
Note that while the resulting $\varepsilon_{12}$
are quite similar in the three approximations in the whole $1S$
regime, and in the $2S$ regime with sizable occupation of the
second subband, noticeable differences arise in the limit of small
second subband occupancy. Starting with the X(X)-C(LDA)
approximation, $\varepsilon_{12}$ shows an exchange-driven abrupt
\emph{positive} jump at the $ 1S \rightarrow 2S$\cite{rigamonti2}.
The inclusion of C(pX) overcomes the X(X)-C(LDA) positive jump,
resulting now in a \emph{negative} jump in $\varepsilon_{12}$,
until it levels with the other results at finite second subband
fillings. LDA is in between the two kind of
discontinuities at $\Gamma = 0$, showing only a discontinuity in
the derivative. The lower inset shows the corresponding
$V_{xc}(z)$ in the relevant QW region. $V_{xc}(z)$ as resulting
from X(X)-C(LDA) builds a barrier just where most of the weight of
$\xi_{2}(z)$ is concentrated, pushing $\varepsilon_{2}$ upwards
under small occupancy of the second subband, and leaving
$\varepsilon_{1}$ more or less unaltered; this explains the
discontinuous positive behavior of $\varepsilon_{12}$. It is also
physically reasonable: by blocking the occupation of the second
subband, the exchange energy is optimized, as the intrasubband
exchange is larger than the intersubband exchange. The $V_{xc}(z)$
in the X(X)-C(pX) behaves just in the opposite way: it develops a
deep well just after the transition, inducing an abrupt decrease
of $\varepsilon_{2}$ at more or less constant $\varepsilon_{1}$.
This explains the abrupt decrease of $\varepsilon_{12}$ in this
case. The behavior has again a simple physical explanation: by
inducing the occupancy of the second subband, $V_{xc}(z)$ promotes
a spatial separation between electrons in both subbands,
decreasing correlation and its associated repulsive energy. With
respect to the $V_{xc}(z)$ resulting from LDA, it shows the
expected smooth and continuous behavior at the transition. 
  
Besides
these fully numerical results, we provide below an analytical
derivation of the results of Fig. 1, for $\Gamma\simeq0$. In the
limit $|\Gamma|\rightarrow0$, $E_{xc}$ can be approximated as
\begin{equation} E_{xc}^{(\alpha)}=P_{xc}^{(\alpha)}+\Gamma
Q_{xc}^{(\alpha)},\label{exc}\end{equation}
 with $\alpha=1S,2S$. Here, $P_{xc}^{(1S)}=E_{xc}(\Gamma\rightarrow0^{-})$,
$P_{xc}^{(2S)}=E_{xc}(\Gamma\rightarrow0^{+})$,
$Q_{xc}^{(1S)}=dE_{xc}/d\Gamma|_{0^{-}}$,
$Q_{xc}^{(2S)}=dE_{xc}/d\Gamma|_{0^{+}}$. By inspection of the
explicit expressions for $E_{x}$ and $E_{c}^{GL(2)}$ given in I,
it is concluded that $P_{xc}^{(1S)}=P_{xc}^{(2S)}=P_{xc}$, and
that $Q_{xc}^{(1S)}\neq Q_{xc}^{(2S)}$. In words, for a fixed set of 
$\varepsilon_{i \ne 2}$'s and $\xi_{i}(z)$'s, the $xc$
functional is continuous at the $1S \rightarrow 2S$, but its
derivative is discontinuous. The explicit expressions for
$P_{xc}$, $Q_{xc}^{(1S)}$ and $Q_{xc}^{(2S)}$ are not needed for
the present derivation. Inserting Eq.(4) in Eq.(3) we obtain,
\begin{equation} V_{xc}^{(\alpha)}(z)\!=\!\!\! \int\!\!\left[\!\frac{\delta
P_{xc}}{\delta
V_{KS}(z')}\!-\!Q_{xc}^{(\alpha)}|\xi_{2}(z')|^{2}\right]\!\chi_{\alpha}^{-1}(z',z)dz'.\label{vxc}\end{equation}
Here, we have defined $\chi_{\alpha}^{-1}(z,z')\equiv\delta V_{KS}(z)/\delta n_{\alpha}(z')$,
and used the result $\delta\Gamma/\delta
V_{KS}(z)=-\delta\varepsilon_{2}/\delta
V_{KS}(z)=-|\xi_{2}(z)|^{2}$, obtained by first-order perturbation theory.
We have also neglected a term lineal in $\Gamma$, which becomes
arbitrarily small in the limit $\Gamma\rightarrow0$. In the limit
$\Gamma\rightarrow0^{+}$, the density response
function\cite{rigamonti1} becomes\begin{equation}
\chi_{2S}(z,z')=\chi_{1S}(z,z')-\left|\xi_{2}(z)\xi_{2}(z')\right|^{2}/\pi.\label{chi}\end{equation}
Its inverse can be calculated by using the Sherman-Morrison
technique\cite{numrec}. We obtain
\begin{equation}
\chi_{2S}^{-1}(z,z')=\chi_{1S}^{-1}(z,z')+D(z)D(z')/[\pi(1+\lambda)],\label{chiinv}\end{equation}
 where $D(z)=\int\chi_{1S}^{-1}(z,x)|\xi_{2}(x)|^{2}dx$, and $\lambda=-\pi^{-1}\int D(x)|\xi_{2}(x)|^{2}dx$.
As we can see from Eq.(7), $\chi_{2S}^{-1}(z,z')$ is discontinuous
at the $1S \rightarrow 2S$ transition, such as it is
$\chi_{2S}(z,z')$ of Eq.(6). Using now Eq.(5) for $\alpha=1S,2S$
and exploiting the explicit expression for $\chi_{2S}^{-1}(z,z')$
as given by Eq.(7), we arrive at the result
%\begin{equation}
%\Delta V_{xc}(z)\equiv V_{xc}^{(2S)}(z)-V_{xc}^{(1S)}(z)=\frac{C_{xc}}{(1+\lambda)}D(z),\label{deltavxc}
%\end{equation}
\begin{equation}
\Delta V_{xc}(z)\equiv V_{xc}^{(2S)}(z)-V_{xc}^{(1S)}(z)=C_{xc}D(z)/(1+\lambda),\label{deltavxc}
\end{equation}
 with $C_{xc}=\langle V_{xc}^{(1S)}\rangle_{2}/\pi-\Delta Q_{xc}$,
$\left\langle {\cal O}\right\rangle _{i}=\int
dx|\xi_{i}(x)|^{2}{\cal O}(x)$ and $\Delta
Q_{xc}=Q_{xc}^{(2S)}-Q_{xc}^{(1S)}$. Eq.(8) is an important
result, which shows explicitly how the functional dependence on
$z$ of the $xc$ potential changes discontinuously at the $1S
\rightarrow 2S$ transition.
For an arbitrary subband transition $NS\rightarrow (N+1)S$ it can be shown 
that the result of Eq.(8) is still valid, with the replacements $1\rightarrow N$ 
and $2\rightarrow N+1$ in Eqs.(4)-(8). 
Equation 8 follows rigorously from Eq.(4), and
then it is important to discuss its validity. In writing Eq.(4) we
have implicitly made a "frozen" assumption, by
taking the same set of wavefunctions $\xi_{i}(z)$ and energies
$\varepsilon_{i}$ both for $\Gamma\rightarrow0^{-}$ and
$\Gamma\rightarrow0^{+}$. In the language of the numerical
self-consistent calculations leading to Fig. 1, it is as if the
results for the $1S$ case in the limit $\Gamma\rightarrow0^{-}$
were extrapolated to the $2S$ $\Gamma\rightarrow0^{+}$ case, by
performing a single iteration loop towards full self-consistency.
This "frozen" result for the $xc$ energy
functional is illustrated in Fig. 2 by the straight lines,
separating exchange ($P_x + \Gamma Q_x^{\alpha}$) from correlation ($P_c +
\Gamma Q_c^{\alpha}$). Besides, if the approximation for $E_{xc}$ is such
that $\Delta V_{xc}(z)\neq0$ (that is, if $C_{xc}\neq0$), it is
clear that the self-consistent iteration loop will lead to a
further discontinuity in the $xc$ energy functional itself right
at the $1S \rightarrow 2S$ transition $(\Gamma=0)$, once
convergence has been reached. These fully self-consistent results
correspond to the thick full ($E_x$) and dashed ($E_c^{GL(2)}$)
lines in Fig. 2. On the other side, if the approximation for
$E_{xc}$ is such that $C_{xc}=0$, no discontinuity of the type of
Eq.(8) exists for $V_{xc}(z)$ before or after self-consistency is
achieved, which in turn implies the continuity of $E_{xc}$. Our
$E_{xc}=E_{xc}[\{\varepsilon_{i}^{\sigma},\xi_{i}^{\sigma}\}]$ is
such that $C_{xc}\neq0$; local (LDA) approximations for $E_{xc}$
yield $C_{xc}=0$.

\begin{figure}
\includegraphics[%
  bb=55bp 249bp 530bp 710bp,
  clip,
  width=8cm,
  keepaspectratio]{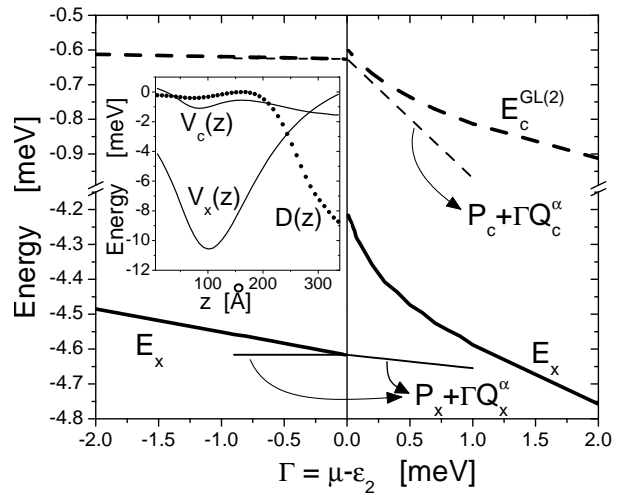}

\caption{Full lines: $E_x$ in the "frozen" (thin
line) and self-consistent (thick line) approaches; dashed lines:
$E_c^{GL(2)}$ in the "frozen" (thin line) and
self-consistent (thick line) approaches. Note the cut in the
vertical scale. Inset: exchange and
correlation potentials (full lines) and function $D(z)$ (dotted line, arbitrary units) for $\Gamma=-\hspace{2pt}0.01$ meV.}
\end{figure}
For further insight, the constant $C_{xc}$ can be separated
conveniently in its exchange and correlation components
$C_{n}=\langle V_{n}^{(1S)}\rangle_{2}/\pi-\Delta Q_{n}$, with
$n=x,c$. The analysis of the results shown in Fig. 2 leads to the
conclusion that $C_{x}<0$ and $C_{c}>0$, considering that $D(z)<0$
(see inset in Fig. 2). Qualitatively, this happens in the following way:
\emph{a)} in the exchange case, $\langle V_{x}^{(1S)}\rangle_{2}$
is large and negative (see inset Fig. 2), while $-\Delta Q_{x}$ is
a relatively small positive magnitude, resulting in $C_{x}<0$;
\emph{b)} in the correlation case, $\langle
V_{c}^{(1S)}\rangle_{2}$ is a very small quantity (see inset Fig.
2), while $-\Delta Q_{c}$ is a relatively large positive number,
yielding a $C_{c}>0$. The net result is that correlation overcomes
exchange ($C_{xc} = C_x + C_c
> 0$), and $\Delta V_{xc}(z)$ has a negative contribution
resulting in the right attractive well shown in the lower inset of
Fig. 1. It is important to note that this overcoming of
correlation on exchange happens even when
$|E_{c}^{GL(2)}|\ll|E_{x}|$ (Fig. 2). However, as Eq.(8) clearly
shows, not only the magnitude of the $xc$ energy functional
matters (represented by the $\langle V_{xc}^{(1S)}\rangle_{2}$
contribution), but also the respective derivatives (represented by
the $\Delta Q_{xc}$ term). 
This exemplifies quite vividly the
potential danger of neglecting correlation against exchange at the 
threshold of a subband transition, under
the argument that the correlation energy is much
smaller than the exchange energy.
 
\begin{figure}
\includegraphics[%
  bb=64bp 325bp 572bp 740bp,
  clip,
  width=8cm,
  keepaspectratio]{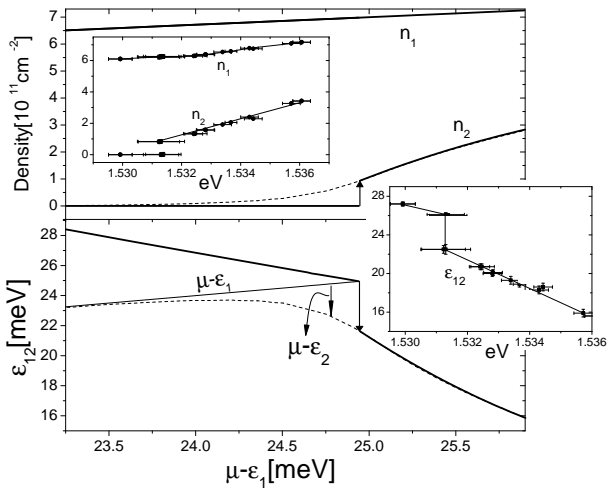}

\caption{Thick full lines: Ground and first-excited subband
densities $n_{1}$ and $n_{2}$ (upper panel) and intersubband
transition energy $\varepsilon_{12}$ (lower panel) as a function
of $\mu-\varepsilon_{1}$. Dashed lines: $n_{2}$ and
$\varepsilon_{12}$ corresponding to a continuous filling of the
first-excited subband, as in Fig. 1.  Insets: experimental data
from Ref.[\onlinecite{goni}].}
\end{figure}

Is there any experimental evidence of this type of discontinuities
in semiconductor QW's? The answer is yes. We reproduce as an inset
in the upper panel of Fig. 3 the experimental values for the
subband densities $n_{1}$ and $n_{2}$, plotted as a function of
the Fermi level measured from the top of the valence
band\cite{goni}. The data have been obtained from a quantitative
analysis of photoluminescence line shapes. It is seen that when
the Fermi level touches the bottom of the second subband, 
the electron density $n_{2}$ jumps from
zero to a finite value ranging from $3\times$ 10$^{10}$ to 8
$\times$ 10$^{10}$ cm$^{-2}$, depending on temperature\cite{goni}.
The electron density of the lowest subband, in contrast, increases
slightly but smoothly with voltage, suggesting that the external
electric field couples essentially to the $1S$ occupation, which
seems quite plausible as it is the subband with the largest
occupation. This motivates us to redraw the results for
$\varepsilon_{12}$ of Fig. 1 as a function of
$\mu-\varepsilon_{1}$, in the lower panel of Fig. 3. Clearly the
theoretical $n_{1}$, $n_{2}$, and $\varepsilon_{12}$ vs.
$\mu-\varepsilon_{1}$ agree quite well with the experimental data,
both qualitatively and quantitatively. For instance, the
theoretical value for the decreasing jump in $\varepsilon_{12}$ at
$\mu-\varepsilon_{1}\simeq$ 25 meV, amounts to about 3.3 meV, in
excellent agreement with the 3.5 meV jump estimated from
experiment.

The results presented in this work are intimately related to the
issue of the derivative discontinuity of ensemble
DFT\cite{perdew1}. Among the many important consequences derived
from this extension of DFT to fractional particle number, maybe
the most important one in the field of solid state physics has
been the realization that the true gap of semiconductors is not
given by the KS one-particle gap\cite{perdew2}. Instead, the true
gap is given by the sum of the KS gap, plus the so-called $xc$
discontinuity, $\Delta_{xc}$. Continuum approximations to $xc$
energy functionals, including all currently widely used LDA's and
GGA's, fail to produce the correct value for $\Delta_{xc}$,
resulting in an important underestimation of the fundamental
band-gap of most semiconductors and insulators. In a very recent
work, Gr\"{u}ning \emph{et al.} have clarified the
theoretical situation, obtaining very good agreement between experimental
and theoretical band gaps of Si, LiF, and Ar\cite{gruning}, by using
an orbital based correlation energy functional corresponding to a dynamical
screening of the Coulomb interaction (GW approximation). It was
found that $\Delta_{xc}$ contributes as much as $30\%-50\%$ to the
energy gap. Their results, for a different type of systems, are fully consistent
with ours.

In conclusion, the intrinsic discontinuity of the $xc$ energy
functional has been obtained entirely within a DFT framework, for
a realistic system. We have shown that the $xc$ energy functional
generated by second-order G\"{o}rling-Levy perturbation theory for the
\emph{2DEG} has many of the properties of the exact functional.
The main finding is that at the $1S \rightarrow 2S$ transition,
the KS $xc$ potential and the associated intersubband transition
energies behave discontinuously, with $x$ and $c$ giving opposite
contributions (i.e. competing) to the discontinuity, and with
correlation overcoming exchange. Very good qualitative and
quantitative agreement is obtained with experiments.

This work was partially supported by CONICET under grant PIP 5254
and the ANPCyT under PICT 03-12742. SR acknowledges financial
support from CNEA-CONICET. CRP is a fellow of CONICET.


\begin{thebibliography}{10}
\bibitem{parr}R. G. Parr and W. Yang, \emph{Density Functional Theory of Atoms and
Molecules}, (Oxford University Press, New York, 1989); R. M. Dreizler
and E. K. U. Gross, \emph{Density Functional Theory} (Springer-Verlag,
Heidelberg, 1990).
\bibitem{tao}J. Tao \emph{et al.}, Phys. Rev. Lett. \textbf{91}, 146401 (2003).
\bibitem{mattsson}A. E. Mattsson, Science \textbf{298}, 759 (2002).
\bibitem{gorling}A. G\"{o}rling and M. Levy, Phys. Rev. B \textbf{47}, 13105 (1993);
\emph{ibid}, Phys. Rev. A \textbf{52}, 4493 (1995).
\bibitem{rigamonti1}S. Rigamonti and C. R. Proetto, Phys. Rev. B \textbf{73}, 235319 (2006).
As the calculations of this work were restricted to the simpler $1S$
regime, \textit{none} of the findings of the present work concerning
the $1S\rightarrow2S$ transition were considered. There is also an important
difference from the technical point of view: the calculation of $E_{c}^{GL(2)}$ in
a situation with two (or more) occupied subbands is far more complicated and 
numerically demanding than the $1S$ case, due to the presence in the many-subband 
case of four Fermi disk intersecting integrals, whose numerical evaluation
becomes quite involved.
\bibitem{mori}P. Mori-S\'{a}nchez, Q. Wu, and W. Yang, J. Chem. Phys. \textbf{123},
062204 (2005).
\bibitem{jiang}H. Jiang and E. Engel, J. Chem. Phys. \textbf{123}, 224102 (2005).
\bibitem{note}As is well
known, $E_{c}^{GL(2)}$ diverges in the long-wavelength limit for
the case of the homogeneous \emph{3D} electron gas\cite{pines}. In
contrast, this contribution is finite for the strict \emph{2D}
homogeneous electron gas\cite{rajagopal}. Our QW system is much
closer to the \emph{2D} than to the \emph{3D} limit (corresponding
to a very large number of occupied subbands). It seems then quite
plausible that the perturbative expansion of Eq.(2) be much better
convergent for the \emph{2DEG} than for the \emph{3DEG}. From a
practical point of view, evidence on this stems from the fact that
the self-consistent loop among Eqs.(1)-(3) is robust and free from
the instabilities found in Refs.\cite{mori} and \cite{jiang}.
\bibitem{rigamonti2}S. Rigamonti, C. R. Proetto, and F. A. Reboredo, Europhys. Lett. \textbf{70},
116 (2005).
\bibitem{numrec}W. H. Press \emph{et al.}, in \emph{Numerical
    Recipes} (Cambridge University Press, NY, 1992).
\bibitem{goni}A. R. Go\~{n}i \emph{et al.}, Phys. Rev. B \textbf{65}, 121313(R)
(2002).
\bibitem{perdew1}J. P. Perdew \emph{et al.}, Phys. Rev. Lett. \textbf{49}, 1691 (1982).
\bibitem{perdew2}J. P. Perdew and M. Levy, Phys. Rev. Lett. \textbf{51}, 1884 (1983);
L. J. Sham and M. Schl\"{u}ter, Phys. Rev. Lett. \textbf{51}, 1888
(1983). Note the similarity between Eq.(12) of the latter work for the semiconductor
band-gap discontinuity, and our Eq.(8) for the discontinuity of the $xc$ potential 
at the $1S\rightarrow2S$ transition.
\bibitem{gruning}M. Gr\"{u}ning, A. Marini and A. Rubio, J. Chem. Phys. \textbf{124}, 154108 (2006).
\bibitem{pines}D. Pines, \emph{Elementary Excitations in Solids} (Benjamin, New York,
1964).
\bibitem{rajagopal}A. K. Rajagopal and J. C. Kimball, Phys. Rev. B \textbf{15}, 2819
(1977); A. Isihara and L. Ioriatti, ibid. \textbf{22}, 214 (1980).\end{thebibliography}
\end{document}